\documentclass[12pt]{article}

\usepackage{scicite}
\usepackage{graphics,color}
\usepackage[latin1]{inputenc}

\usepackage{times}
\usepackage{amssymb}

\newenvironment{sciabstract}{%
\begin{quote} \bf}
{\end{quote}}

\newcounter{lastnote}

\enlargethispage*{1cm}

\title{Relativistic Frequency Synthesis of Light Fields} 

\author
{C. Rödel,$^{1,2,\ast,\dagger}$ E. Eckner,$^{1,\dagger}$ J. Bierbach,$^{1,2}$ M. Yeung,$^{3}$ B. Dromey,$^{3}$ T. Hahn,$^{4}$ S. Fuchs,$^{1,2}$\\
A. Galestian Pour,$^{1}$ M. Wünsche,$^{1}$ S. Kuschel,$^{1,2}$ D. Hemmers,$^{4}$ O. Jäckel,$^{1,2}$ \\ G. Pretzler,$^{4}$ M. Zepf,$^{2,3}$ G. G. Paulus$^{1,2}$\\
\\
\normalsize{$^{1}$ Institute of Optics and Quantum Electronics, Friedrich-Schiller-University Jena, Germany}\\
\normalsize{$^{2}$ Helmholtz Institute Jena, Germany}\\
\normalsize{$^{3}$ Centre for Plasma Physics, Queen's University Belfast, United Kingdom}\\
\normalsize{$^{4}$ Institute for Laser and Plasma Physics, University of Düsseldorf, Germany}\\
\normalsize{$^\ast$ To whom correspondence should be addressed; E-mail: christian.roedel@uni-jena.de}\\
\normalsize{$^\dagger$ These authors contributed equally.}
}

\date{}

\begin{document} 


\baselineskip24pt


\maketitle 


\begin{sciabstract}
Waveform shaping and frequency synthesis based on waveform modulation is ubiquitous in electronics, telecommunication technology, and optics  \cite{Kundu2013, Lee2002}. For optical waveforms, the carrier frequency $\omega_L$ is on the order of several hundred THz, while the modulation frequencies used in conventional devices like electro- or acousto-optical modulators are orders of magnitude lower \cite{Lee2002}. As a consequence, any new frequencies are typically very close to the fundamental. The synthesis of new frequencies in the extreme ultraviolet (XUV), e.g. by using relativistic oscillating mirrors \cite{Lichters1996, Tsakiris2006}, requires modulation frequencies in the optical regime \cite{Dromey2006, Thaury2007} or even in the extreme ultraviolet. The latter has not been proven possible to date. Here we demonstrate that individual strong harmonics can indeed be generated by reflecting light off a plasma surface that oscillates at XUV frequencies. The strong harmonics are explained by nonlinear frequency mixing of near-infrared light and a laser-driven plasma oscillation in the extreme ultra-violet mediated by a relativistic non-linearity.
\end{sciabstract}

% In setting up this template for *Science* papers, we've used both
% the \section* command and the \paragraph* command for topical
% divisions.  Which you use will of course depend on the type of paper
% you're writing.  Review Articles tend to have displayed headings, for
% which \section* is more appropriate; Research Articles, when they have
% formal topical divisions at all, tend to signal them with bold text
% that runs into the paragraph, for which \paragraph* is the right
% choice.  Either way, use the asterisk (*) modifier, as shown, to

%\section*{Introduction}

\newpage

The modulation of a sinusoidal waveform $A\sin(\omega_L t + \phi)$
 can be regarded as the fundamental building block for generating new frequencies of some fundamental frequency $\omega_L$. A waveform modulation can be achieved by making any of the parameters $A$, $\omega_L$, and $\phi$ dependent on time. Accordingly, amplitude, frequency, and phase modulation are distinguished. For a constant modulation frequency $\omega_m$, for instance, sideband frequencies $\omega_L \pm k\omega_m$ appear, where $k$ are integers.

\begin{figure}[htb]
	\centering
		\includegraphics{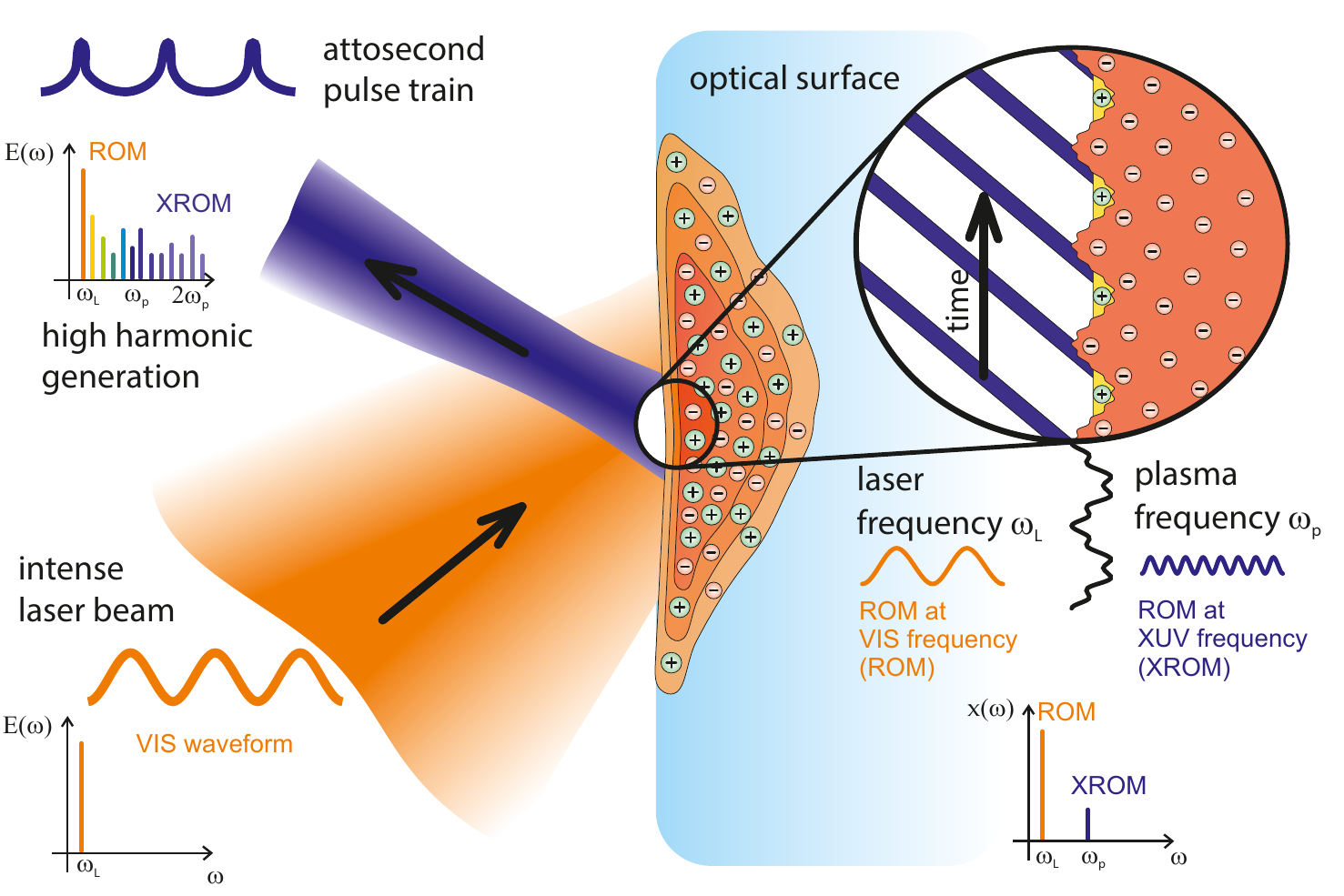}
	\caption{\textbf{Relativistic frequency synthesis of light fields:} An intense laser pulse is focused onto a surface where a plasma is generated which reflects the laser beam. For high laser intensities $> 10^{18} \, \rm W/cm^2$, the electron density at the surface oscillates relativistically while the ion background can be regarded as fixed. Thus, a high harmonic spectrum is generated in the reflected beam when the plasma is moving normal to the surface as described by the model of a Relativistic Oscillating Mirror (ROM). The motion spectrum $x(\omega)$ of the ROM typically consists of the excitation at optical frequencies - most effectively driven at the laser frequency $\omega_L$. In addition, the dense surface plasma is oscillating at the plasma frequency $\omega_p$ in the XUV which leads to a nonlinear frequency mixing in the XUV (XROM) and the enhancement of particular harmonic frequencies around $2\omega_p$.}
	\label{ROM}
\end{figure}

Phase modulation of light upon reflection from relativistically oscillating plasma surfaces has been proposed as a mechanism for efficient high-harmonic generation (HHG). For reviews see Ref.~\cite{Quere2010} \& \cite{Teubner2009}. In this process, a dense plasma from a solid density surface located at $x(t)$ oscillates such that relativistic velocities $\beta=v/c \lesssim 1$ are reached. At times when the relativistic Lorentz factor $\gamma(t)=1/\sqrt{1-\beta^2}$ is large and the plasma is moving towards the incident light wave with high velocity $\beta^{\rm max}$, high frequency radiation is generated in the reflected beam by the relativistic Doppler effect \cite{Gordienko2004,Quere2010}. Due to this mirror-like reflection with a periodicity of $\omega_L$, this process can be interpreted as a phase modulation \cite{Bulanov1994, Lichters1996, Tsakiris2006}. Consequently, it is often referred to as the Relativistic Oscillating Mirror (ROM), see Fig. 1. Note that the relativistic plasma oscillation $x(t)=\frac{\beta^{\rm max}_L}{\omega_L} \sin(\omega_L t+\varphi_L)$ discussed in the literature to date is at the laser frequency $\omega_L$ or its low order harmonics $n \omega_L$ \cite{Lichters1996,Tsakiris2006}, i.\,e. at optical frequencies. Models that include the relativistic effect of retardation $t=t'-x(t')/c$ show good agreement with numerical simulations \cite{Lichters1996,Quere2010} by accounting for the difference of the time of reflection $t'$ and the time of observation $t$. In the limit of a strongly relativistic motion such models predict a harmonic spectral envelope spanning a broad bandwidth in the XUV and soft x-ray range that follows a power law $(\omega/\omega_L)^{-8/3}$ up to some roll-off frequency $\omega_{\rm ro}\propto \gamma^3$ \cite{Baeva2006} after which the harmonic intensity decreases exponentially. Moreover, the shallow spectral slope not only implies that coherent high frequency radiation can be efficiently generated by this harmonic generation mechanism, but rather is also a signature of a train of extremely short optical pulses in the time domain \cite{Baeva2006, Quere2010, Behmke2011}. HHG from relativistic surfaces thus is one of the most promising routes for intense attosecond pulse generation \cite{Tsakiris2006, Sansone2011}.

In the last decade, relativistic surface high harmonic generation has been demonstrated up to keV photon energies \cite{Dromey2006} in experiments which have also verified the shallow harmonic slope and the scaling of $\omega_{\rm ro}$ \cite{Dromey2007}. At more moderate intensities of the order of $10^{19} \rm \, W/cm^2$, which are available at many laboratories, relativistic surface harmonic radiation has been generated for different laser and plasma parameters \cite{Thaury2007, Tarasevitch2007, Krushelnick2008}. Recently, efficiencies ranging from $10^{-4}$ - $10^{-6}$ with a strong dependence on the plasma gradient $L_p$ have been reported and generally reveal a decrease of the harmonics efficiency for very short scale lengths $L_p \ll \lambda/10 $ \cite{Rodel2012, Kahaly2013}. For short gradients $L_p$ of the order of $\lambda/10$, i.e. for plasma conditions which can be considered to be close to ideal for efficient generation, a constant, nearly diffraction limited beam divergence has been found for ROM harmonics \cite{Dromey2009}. All experiments so far report a decaying spectral envelope towards higher harmonic orders, possibly with small modulations spanning a few harmonic orders \cite{Watts2002, Teubner2003, Behmke2011} which are explained within the existing theoretical framework of the ROM mechanism where the surface oscillates only at optical frequencies.  

\begin{figure}[htb]
	\centering
		\includegraphics{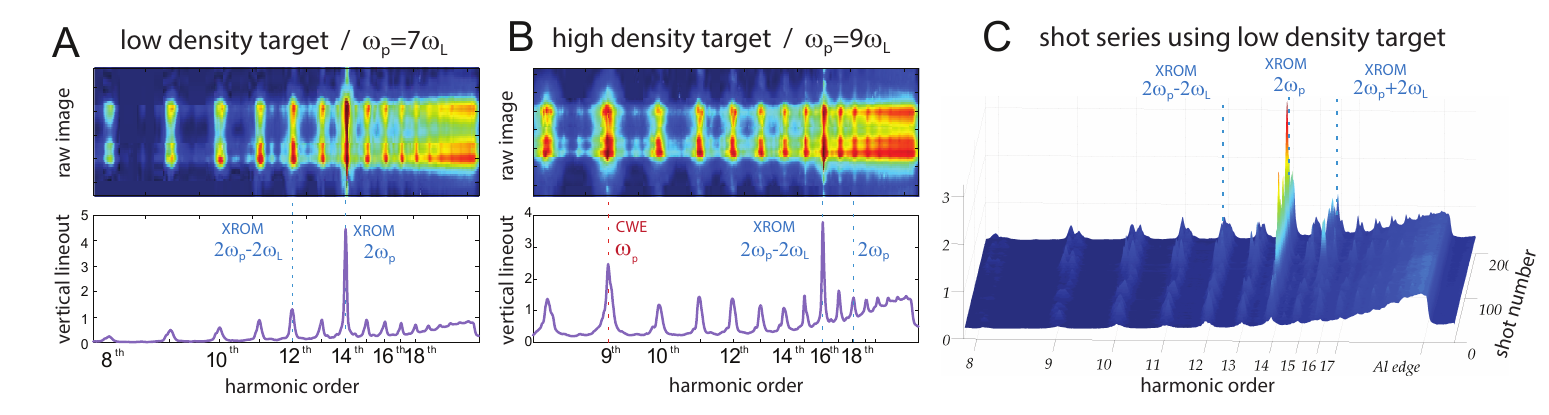}
	\caption{\textbf{High harmonic spectra for different target densities} (upper graph: raw data from CCD camera, lower graph: vertical lineout) \textbf{A:} The high harmonic spectrum using a plastic coated glass surface as a target material ($\omega_p \simeq 7 \omega_L$) shows a strongly increased $14^{th}$ harmonic at $\simeq 2\omega_p$.  \textbf{B:} Using a surface of fused silica the plasma density the plasma frequency is increased to $\omega_p \approx 9 \omega_L$ (CWE harmonic at $\omega_p$). The enhanced harmonic appears at $2\omega_p-2\omega_L$. \textbf{C:} Using the plastic surface 200 subsequent harmonic spectra are recorded while the laser intensity is fluctuating at moderate relativistic intensities ($10^{18}-10^{19} \, \rm W/cm^2$). The harmonic intensity (linear scale) at $2\omega_p$ and unsteadily at $2\omega_p+2\omega_L$ is increased over the entire data set. For nonrelativistic intensities ($<10^{18} \, \rm W/cm^2$) the enhanced harmonic emission near $2\omega_p$ vanishes.}
	\label{spectra}
\end{figure}

Here we report on the strongly enhanced emission of particular harmonic orders under conditions where the plasma density profile is approaching a step function. Depending on the target density, either the 14th (43.4\,eV), 16th (49.6\,eV), or 18th (55.8\,eV) harmonic of a 400-nm driving terawatt laser pulse has been amplified by a factor of up to five as compared to the adjacent harmonics (see Fig.~\ref{spectra}A/B). Moreover, the enhancement is never observed for odd harmonic orders or nonrelativistic intensities. In addition, the enhancement can also appear at side bands of an amplified harmonic -- but again only at even harmonic orders. We note that the spectral width of an enhanced harmonic is typically very similar to that of the neighboring harmonics (Details in S~5) which are generated by the familiar ROM process. The enhanced harmonics contain a pulse energy of $\approx 1 \mu J$ and are thus already suited for applications like coherent diffraction imaging \cite{Sandberg2007} or seeding of free electron lasers \cite{Lambert2008}. We further note that the generation efficiency of the enhanced harmonics is more than one order of magnitude higher than the ROM harmonics in this frequency range which are obtained from plasma density scale lengths $L_p \simeq \lambda/5$ that have been considered to be ideal so far \cite{Rodel2012,Kahaly2013}. In the following we show that the enhanced harmonics are generated by a relativistic plasma surface oscillating at XUV frequencies and, therefore, refer to the high-frequency XUV ROM as XROM.

A detailed evaluation of a set of harmonic spectra (Fig. \ref{spectra}C) reveals that fluctuations in the intensity of the 400-nm fundamental intensity do \textit{not} affect the order of the enhanced harmonic. However, the order of the enhanced harmonic is affected by the target material. For plastic targets (low density), it is the $14^{th}$ harmonic (28.6\,nm), for glass targets (high density) either the $16^{th}$ or $18^{th}$. The enhanced harmonics are in fact conspicuously close to twice of the maximum plasma frequency $\omega_p=\sqrt{n_e e^2/(\varepsilon_0 m_e)}$ which is the natural frequency for electrons in a plasma with an electron density $n_e$. Here, $e$ is the charge of an electron, $\varepsilon_0$ is the vacuum permittivity and $m_e$ is the electron mass. The value of $\omega_p$ was verified to be near the $7^{th}$ harmonic for plastic targets or near the $9^{th}$ harmonic for glass targets, respectively, by measuring the cutoff frequency for harmonics  at lower, non-relativistic intensities. At nonrelativistic conditions it is well established that Coherent Wake Emission (CWE) \cite{Quere2006} becomes dominant and that the high harmonic spectrum ends at $\omega_p$ \cite{Thaury2007}. It is worth noting that the CWE is strongest for the cutoff harmonic located at $\omega_p$, e.g. the $9^{th}$ harmonic in Fig.~\ref{spectra}B, which is a signature of an extremely steep plasma density ramp \cite{Dromey2009a} (Details in S~3). However, at this frequency range the complex superposition of harmonics generated by different generation mechanisms makes it nearly impossible to assign the harmonics around $\omega_p$ to CWE, the ROM or -- as we will see later -- to the XROM only by experimental data.

Before explaining the physical details of the XROM, we briefly review the observation of radiation emitted at the plasma frequency and its multiples. Plasma radiation at $\omega_p$ and $2\omega_p$ is a well-known observation in gas discharges, tokamak plasmas \cite{Fidone1978} and astrophysical plasmas such as auroral \cite{Forsyth1949} and solar radio burst \cite{Wild1953} and is, in general, incoherent. In laser produced plasmas, radiation at $\omega_p$ and $2\omega_p$ \cite{Teubner1999} has been observed which has been attributed to the coupling of two large-amplitude Langmuir waves and subsequent two-plasmon decay (TPD) \cite{Boyd2000, Kunzl2003}. Common to this type of incoherent, broadband plasma radiation is the requirement of plasma waves generated inside the plasma and, in general, their nonrelativistic origin. Thus their appearance should not be strongly intensity dependent and their spatial and spectral properties would not be linked to ROM harmonic orders. Conversely, the similarity of ROM- and XROM-harmonics in terms of divergence, intensity dependence and bandwidth indicates that both spatial and temporal phase are comparable to the ROM mechanism and hence suggest strongly that the origin of XROM harmonics is not coupled to TPD or CWE but rather to a ROM-like mechanism.

\begin{figure}[htb]
	\centering
		\includegraphics{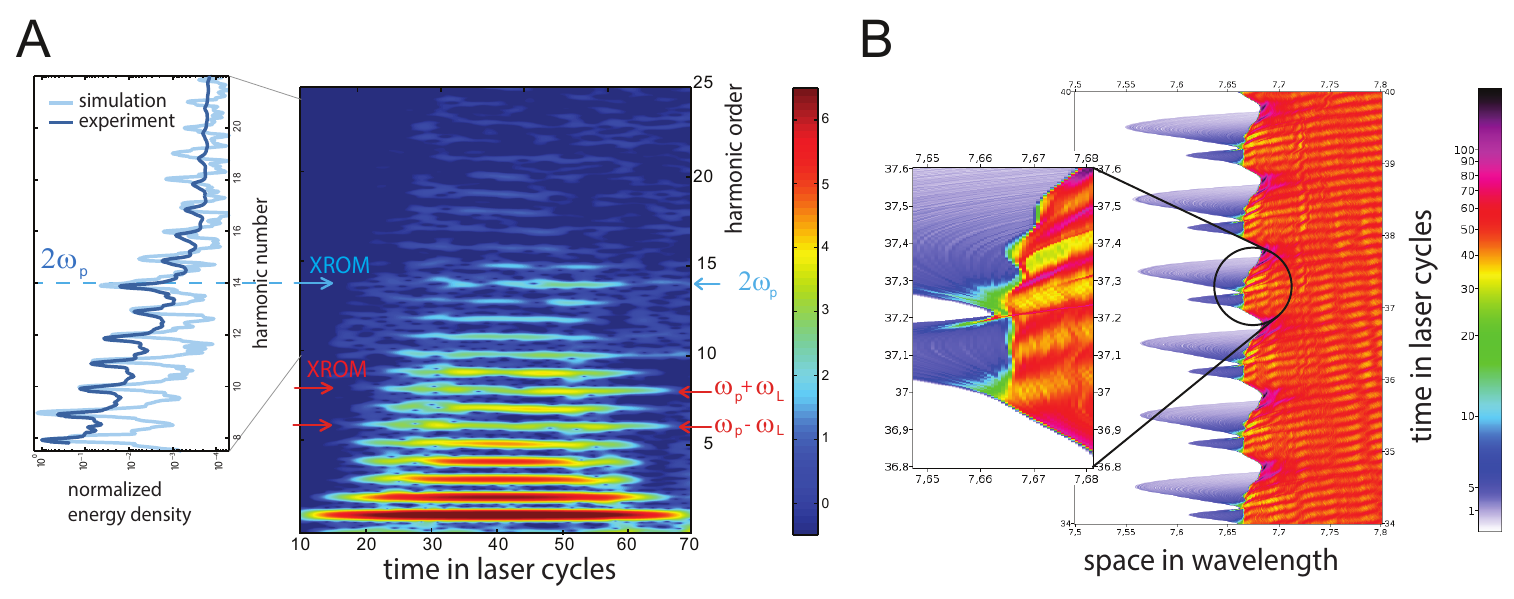}
	\caption{\textbf{Laser plasma simulation A:} A laser plasma simulation is performed using the parameters of the experiment with the low density target ($a_0=2.5, \, n_e=49n_c,\, \omega_p=7\omega_L$). The reflected field is analyzed using a time-windowed Fourier transform. The harmonic intensity (logarithmic scale) at $2\omega_p$ is increased as compared to their adjacent harmonics thus reproducing our experimental result. In addition, the XROM harmonic at $2\omega_p$ is emitted for a slightly longer time than the other harmonics in this frequency range - an effect which is also observed at the sideband harmonics of the plasma frequency located at $\omega_p \pm \omega_L$. It is shown that these harmonics are also increased by the XROM mechanism. 
	  \textbf{B:} The electron density as a function of time is displayed showing that the plasma is oscillating with different frequencies. Most prominently, the surface plasma  is oscillating with $\omega_L$ and, in addition, with the plasma frequency $\omega_p$ (zoom in). Jets of Brunel electrons (blue) are excited mainly with $\omega_L$ into the dense plasma where they excite Langmuir waves at $\omega_p$.}
	\label{PIC}
\end{figure}

Particle-in-cell (PIC) simulations were performed to elucidate the origin of the enhanced harmonics. For simulation parameters that closely match our experimental conditions, the temporal evolution of the electron density at the plasma surface is shown in Fig.~\ref{PIC}B and clearly reveals an oscillation not only at the optical frequency $\omega_L$ of the 400-nm laser but also a strong excitation at the plasma frequency $\omega_p$ in the XUV \cite{comment}. Although the amplitude of this oscillation $a_p=\beta^{\rm max}_p/\omega_p$ is an order of magnitude smaller than the surface oscillation at $\omega_L$, its peak velocity $\beta^{\rm max}_p \approx 0.1 c$ is comparable and can therefore also be expected to result in relativistic nonlinear effects in the reflected spectrum. The mechanism for the excitation of the strong plasma surface oscillation at $\omega_p$ is likely due to the jets of electrons which are expelled to the vacuum side by the laser field and which are later reinjected into the dense plasma \cite{Brunel1987, Geindre2010} (see Fig.\ref{PIC}B). The reflected field of the simulation is analyzed by a time-windowed Fourier transform which is displayed in Fig.~\ref{PIC}A. The time integrated spectrum shows the enhanced harmonic at $2\omega_p$ on top of the familiar spectral decay of the harmonics and thus reproduces the respective experimental result. Further, we note that the $\omega_p$ surface mode seems to affect the emission of the harmonics at $2\omega_p$ and $\omega_p \pm \omega_L$. The strongest emission of the enhanced harmonic at $2\omega_p$ can be found in the second half of the laser pulse. In fact, a similar effect can be seen at the ROM harmonics at $\omega_p \pm \omega_L$, i.e. the sideband frequencies of the $\omega_p$ oscillation. This delayed emission of the enhanced harmonics suggests that the relativistic surface plasma mode first needs to grow over several laser cycles or a few tens of femtoseconds and is damped again when the laser field ceases. Hence, when the relativistic surface plasma mode is pronounced and locked at a multiple integer of the laser frequency the harmonics enhanced by the XROM mechanism are generated. The XROM enhanced harmonics have a pulse duration comparable to ROM harmonics driven by the 45-fs laser pulses. This explains why the XROM harmonics have a similar bandwidth as compared to the adjacent harmonics. Moreover, a scan of the pulse duration in the simulations shows that the enhanced harmonics are observed in particular for pulse durations of a few tens of femtoseconds ($\gtrsim 20 \,$fs).

\begin{figure}[htb]
	\centering
		\includegraphics{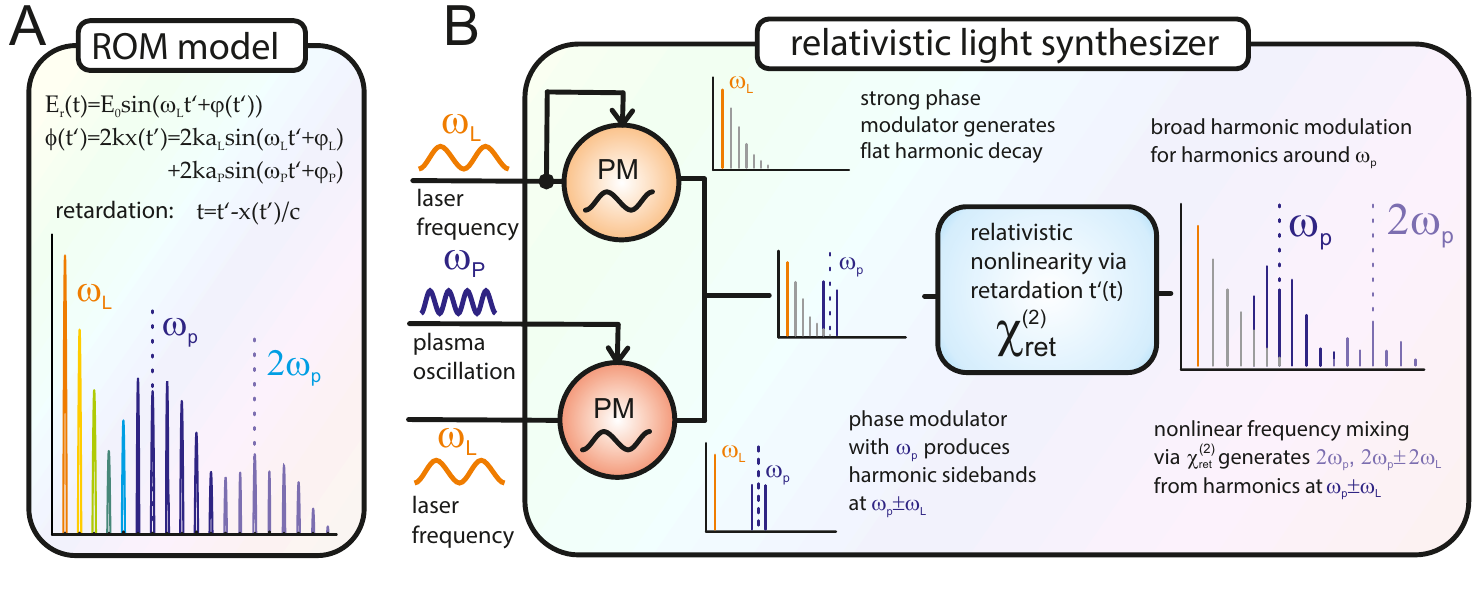}
	\caption{\textbf{Relativistic synthesizer A:} The ROM model including retardation explains the observed harmonic spectra. The low order harmonics have a decay owing to the phase modulation with $\omega_L$. The phase modulation with $\omega_p$ generates strong sideband frequencies $\omega_p \pm \omega_L$ which are accompanied by a harmonic decay similar to that of the fundamental $\omega_L$. When retardation is included, additional harmonics near $2\omega_p$ appear.
\textbf{B:} The harmonic spectrum given by the ROM model can be interpreted by the frequency synthesis of two phase modulators with $\omega_L$ and $\omega_p$. When only considering the second order nonlinearity $\chi^{(2)}_{ret}$ induced by the retardation, the harmonics at $2\omega_p$ and $2\omega_p \pm 2\omega_L$ are enhanced due to sum frequency generation (SFG) and second harmonic generation (SHG) of the enhanced sideband harmonics at $\omega_p \pm \omega_L$.}
	\label{synthesizer}
\end{figure}

Using the approach detailed in Ref.~\cite{Quere2010}, it can be shown that the XROM harmonics are indeed due to a Doppler-upshift from an oscillating mirror. The Fourier transform of current density extracted from the PIC data provides a clear signature of Doppler-type processes with which the competing processes can be distinguished (Details in S\,7). In order to determine which surface velocity components are dominant, an analytic model based on the ROM model was developed. This model explains the observed spectra in a semi-quantitative way by considering the phase modulation of light reflected at a mirror oscillating relativistically with $\omega_L$ {\em and} an additional high frequency modulation $x(t')\propto \sin(\omega_p t' + \phi_{\omega_p})$ (Fig. \ref{synthesizer}). This means that the reflected field $E_{r}(t)\propto \sin(\omega_L t + \phi(t'))$ undergoes a phase modulation with the plasma frequency $\omega_p$. The first signature of the additional modulation to be expected are sidebands at $\omega_p \pm n\cdot\omega_L$. Experimentally, the corresponding enhanced harmonics have not been observable because the harmonic spectrum up to $\omega_p$ is dominated by harmonics generated by the CWE mechanism. Furthermore, without adding relativistic retardation in the model, the strong feature around $2\omega_p$ observed in experiment and PIC simulations cannot be reproduced. However, when the effect of relativistic retardation is accounted for correctly, a significant enhancement of harmonics at $2\omega_p$ is obtained demonstrating that retardation provides the major contribution to the phase modulation. When relativistic retardation is included in the model we obtain a set of implicit equations which requires the reflected field to be solved numerically. The parameters were chosen according to the result of the PIC simulations (cf. Fig.\ref{PIC}B). A spectral analysis of the reflected field is shown in Fig.\ref{synthesizer}A on a logarithmic scale. The phase modulation with $\omega_p$ leads to strong harmonic sidebands at $\omega_p \pm n \omega_L$ (red) which decay exponentially, as in the case without retardation. When retardation is included, additional enhanced harmonics at $2\omega_p$ and even sidebands (blue) appear which reproduces our experimental observation.

In order to describe the effect phenomenologically as in conventional nonlinear optics, one can treat the nonlinearity induced by retardation  -- to first order -- by a quadratic susceptibility $\chi_{\rm ret}^{(2)}$. The second order polarization in time domain $P^{(2)}(t)$ can then be written as $P^{(2)}(t)=\varepsilon_0 \int_0^{\infty} \int_0^{\infty} \tilde{\chi}_{\rm ret}^{(2)}(\tau_1,\tau_2)E_{\rm PM}(t-\tau_1)E_{\rm PM}(t-\tau_2) d\tau_1 d\tau_2 \propto E_{\rm XROM}(t)$, where $\tilde{\chi}_{\rm ret}^{(2)}$ is the nonlinear response function induced by retardation and $E_{\rm PM}$ denotes the electric field which is obtained by the pure phase modulation as defined in the introduction. The nonlinear polarization in frequency domain is given by $P^{(2)}(\omega)=\varepsilon_0 \chi^{(2)}_{\rm ret}(\omega; \omega_1,\omega_2) E_{\rm PM}(\omega_1)E_{\rm PM}(\omega_2) \propto E_{\rm XROM}(\omega)$. $\chi_{\rm ret}^{(2)}(\omega;\omega_1,\omega_2)=\int_{0}^{\infty} \int_{0}^{\infty} \tilde{\chi}^{(2)}(\tau_1,\tau_2) e^{i(\omega_1 \tau_1 + \omega_2 \tau_2)} d\tau_1 d\tau_2$ is the nonlinear susceptibility and $\omega=\omega_1 + \omega_2$. This means that sum frequency generation (SFG) or second harmonic generation (SHG) can be achieved via the relativistic nonlinearity $\chi_{\rm ret}^{(2)}$ from the phase-modulated fields $E_{\rm PM}$. Since we know that $E_{\rm PM}$ mainly consists of the two sideband frequencies $\omega_p \pm \omega_L$, the XROM enhanced harmonics at $2\omega_p$ and $2\omega_p \pm 2\omega_L$ can be interpreted as SFG and SHG in this framework. Interestingly in this perspective, it is the effect of retardation - a property of space-time and the finite value of the speed of light -  which induces the decisive contribution to the frequency synthesis by the XROM process.

In conclusion, we found a strong enhancement of harmonic frequencies generated from a plasma surface oscillating at XUV frequencies with relativistic velocity. The enhanced harmonic generation is explained by a nonlinear frequency synthesis mediated by the relativistic effect of retardation. It has to be regarded as the first demonstration of relativistic frequency mixing in the XUV and thus opens a new field of nonlinear optics in the XUV. The nonlinear response of relativistic plasma surfaces at XUV frequencies may be exploited in the future, for instance, as a novel diagnostics for solid density plasmas. The XROM enhanced harmonic radiation may find applications in various scientific fields where intense, coherent XUV radiation of small bandwidth is needed. Examples are the investigation of plasmas in fusion-related research or in laboratory astrophysics since the radiation at $2\omega_p$ is transmitted. Other applications such as XUV spectroscopy, XUV microscopy, coherent diffraction imaging, or the seeding of free-electron lasers are evident.

\section*{Methods}
\textbf{Experiment.} The surface harmonics are generated by focusing 400-nm laser pulses with a pulse duration of 45\,fs and a pulse energy of 100 mJ onto rotating plastic or glass targets such that a fresh interaction surface is provided for each laser pulse. The 400-nm radiation is created by a table-top 40-TW laser system operating at 800\,nm via second harmonic generation (SHG) in a KDP crystal (0.7\,mm). 
\\
Using an off-axis parabolic mirror, the 400-nm pulses are focused to intensities of $I=2 \cdot 10^{19}\rm\,W/cm^2$ (FWHM) which correspond to a normalized vector potential \\ $a_0= \sqrt{I \lambda^2 / (1.37 \cdot 10^{18} \rm \, W/cm^2 \, \mu m^2)}=2$. An angle of incidence of $45^\circ$ and laser polarization parallel to the plane of incidence is chosen. For lower laser intensities ($< 10^{18} \, \rm W/cm^2$), where relativistic effects can virtually be neglected, the XROM harmonics disappear. The intimate link between the enhanced harmonics and the ROM process can also be inferred from the divergence of the harmonic radiation measured with an angle-resolving XUV spectrometer. It shows that the divergence is identical for harmonics close to the enhanced harmonic ($\approx$23\,mrad at $1/e^2$), which is one of the hallmarks of the ROM process \cite{Dromey2009}. 
\\
Due to the quadratic proportionality of SHG on the 800-nm intensity, prepulses of our laser are suppressed much more efficiently \cite{Bierbach2012} than in other experiments using, for instance, a plasma mirror setup for contrast enhancement. This extremely high pulse contrast is essential for realizing surface plasmas with a nearly step-like density gradient. The plasma density gradient was modelled using the hydrodynamical code MULTI-FS. A very short plasma scale length $L_p \simeq \lambda/50$ was found for the estimated SHG pulse contrast (Details in S~2) which can be regarded as a nearly step-like density profile.

\textbf{Simulations.} In the simulations using the LPIC code \cite{Lichters1996}, a 10 $\rm \mu m$ plasma slab with a plasma frequency $\omega_p$ and a step-like density profile was irradiated by a 45-fs, Gaussian-shaped laser pulse with a maximum amplitude $a_0=2.5$ while the ions were kept fixed. The reflected field was recorded as a function of time at the vacuum side of the simulation box. The time window of the analyzed field and the width of the plasma slab was chosen such that reflections of fields and particles at the box edges could not influence the reflected field. The use of thick plasma slab also prevents the formation of two counter-propagating plasma waves of frequency $\omega_p$ during the temporal window of interest and thus suppresses the emission at $2\omega_p$ by two-plasmon decay \cite{Kunzl2003} close to the XROM emission frequency which would otherwise interfere with the interpretation of the simulation. 

\section*{Author contributions}

The experiments were planned, designed and performed by C.R., J.B., M.Y., B.D., T.H., S.F., A.G.P., M.W., S.K. with support from D.H. and O.J. The experimental data were analyzed by C.R., E.E., J.B., M.Z. and GGP.  Simulations were performed by E.E., J.B. and D.H. for the interpretation of the experimental results. The theoretical model has been developped by E.E., C.R. and G.G.P. All authors contributed to the manuscript. M.Z., G.P. and G.G.P. supervised the project and acquired funding.

%\begin{figure}[htb]
%	\centering
%		\includegraphics{bandwidth.pdf}
%	\caption{10 Hz, harmonic bandwidth}
%\end{figure}
%
%\begin{figure}
%	\centering
%		\includegraphics{tfa2.pdf}
%		\caption{time frequency analysis}
%	\label{fig:tfa2}
%\end{figure}

%\vspace{1cm}

%Science guidelines:\\
%Research Paper - 4500 words\\
%Reports - 2500 words\\


\begin{thebibliography}{10}

\bibitem{Kundu2013}
Sudakshina Kundu.
\newblock {\em Analog and Digital Communications}.
\newblock Pearson Education India, 2013.

\bibitem{Lee2002}
M.~Lee, H.~E. Katz, C.~Erben, D.~M. Gill, P.~Gopalan, J.~D. Heber, and D.~J.
  McGee.
\newblock Broadband modulation of light by using an electro-optic polymer.
\newblock {\em Science}, 298(5597):1401--1403, 2002.

\bibitem{Lichters1996}
R.~Lichters, J.~Meyer-ter-Vehn, and A.~Pukhov.
\newblock Short-pulse laser harmonics from oscillating plasma surfaces driven
  at relativistic intensity.
\newblock {\em Physics of Plasmas}, 3(9):3425--3437, 1996.

\bibitem{Tsakiris2006}
G.~D. Tsakiris, K.~Eidmann, J.~Meyer-ter-Vehn, and F.~Krausz.
\newblock Route to intense single attosecond pulses.
\newblock {\em New Journal of Physics}, 8, 2006.

\bibitem{Dromey2006}
B.~Dromey, M.~Zepf, A.~Gopal, K.~Lancaster, M.~S. Wei, K.~Krushelnick,
  M.~Tatarakis, N.~Vakakis, S.~Moustaizis, R.~Kodama, M.~Tampo, C.~Stoeckl,
  R.~Clarke, H.~Habara, D.~Neely, S.~Karsch, and P.~Norreys.
\newblock High harmonic generation in the relativistic limit.
\newblock {\em Nature Physics}, 2(7):456--459, 2006.

\bibitem{Thaury2007}
C.~Thaury, F.~Quéré, J.~P. Geindre, A.~Levy, T.~Ceccotti, P.~Monot,
  M.~Bougeard, F.~Reau, P.~D'Oliveira, P.~Audebert, R.~Marjoribanks, and P.~H.
  Martin.
\newblock Plasma mirrors for ultrahigh-intensity optics.
\newblock {\em Nature Physics}, 3(6):424--429, 2007.

\bibitem{Quere2010}
C.~Thaury and F.~Quéré.
\newblock High-order harmonic and attosecond pulse generation on plasma mirrors: basic mechanisms.
\newblock {\em Journal of Physics B-Atomic Molecular and Optical Physics},
  43(21), 2010.

\bibitem{Teubner2009}
U.~Teubner and P.~Gibbon.
\newblock High-order harmonics from laser-irradiated plasma surfaces.
\newblock {\em Reviews of Modern Physics}, 81(2):445--479, 2009.

\bibitem{Gordienko2004}
S.~Gordienko, A.~Pukhov, O.~Shorokhov, and T.~Baeva.
\newblock Relativistic doppler effect: Universal spectra and zeptosecond
  pulses.
\newblock {\em Physical Review Letters}, 93(11):115002, September 2004.

\bibitem{Bulanov1994}
S.~V. Bulanov, N.~M. Naumova, and F.~Pegoraro.
\newblock Interaction of an ultrashort, relativistically strong laser-pulse
  with an overdense plasma.
\newblock {\em Physics of Plasmas}, 1(3):745--757, 1994.

\bibitem{Baeva2006}
T.~Baeva, S.~Gordienko, and A.~Pukhov.
\newblock Theory of high-order harmonic generation in relativistic laser
  interaction with overdense plasma.
\newblock {\em Physical Review E}, 74(4), 2006.

\bibitem{Behmke2011}
M.~Behmke, D.~an~der Brügge, C.~Rödel, M.~Cerchez, D.~Hemmers, M.~Heyer,
  O.~Jäckel, M.~Kübel, G.~G. Paulus, G.~Pretzler, A.~Pukhov, M.~Toncian,
  T.~Toncian, and O.~Willi.
\newblock Controlling the spacing of attosecond pulse trains from relativistic
  surface plasmas.
\newblock {\em Physical Review Letters}, 106(18), 2011.

\bibitem{Sansone2011}
G.~Sansone, L.~Poletto, and M.~Nisoli.
\newblock High-energy attosecond light sources.
\newblock {\em Nature Photonics}, 5(11):656--664, 2011.

\bibitem{Dromey2007}
B.~Dromey, S.~Kar, C.~Bellei, D.~C. Carroll, R.~J. Clarke, J.~S. Green,
  S.~Kneip, K.~Markey, S.~R. Nagel, P.~T. Simpson, L.~Willingale, P.~McKenna,
  D.~Neely, Z.~Najmudin, K.~Krushelnick, P.~A. Norreys, and M.~Zepf.
\newblock Bright multi-kev harmonic generation from relativistically
  oscillating plasma surfaces.
\newblock {\em Physical Review Letters}, 99(8), 2007.

\bibitem{Tarasevitch2007}
A.~Tarasevitch, K.~Lobov, C.~Wünsche, and D.~von~der Linde.
\newblock Transition to the relativistic regime in high order harmonic
  generation.
\newblock {\em Physical Review Letters}, 98(10), 2007.

\bibitem{Krushelnick2008}
K.~Krushelnick, W.~Rozmus, U.~Wagner, F.~N. Beg, S.~G. Bochkarev, E.~L. Clark,
  A.~E. Dangor, R.~G. Evans, A.~Gopal, H.~Habara, S.~P.~D. Mangles, P.~A.
  Norreys, A.~P.~L. Robinson, M.~Tatarakis, M.~S. Wei, and M.~Zepf.
\newblock Effect of relativistic plasma on extreme-ultraviolet harmonic
  emission from intense laser-matter interactions.
\newblock {\em Physical Review Letters}, 100(12), 2008.

\bibitem{Rodel2012}
C.~Rödel, D.~an~der Brügge, J.~Bierbach, M.~Yeung, T.~Hahn, B.~Dromey,
  S.~Herzer, S.~Fuchs, A.~Galestian Pour, E.~Eckner, M.~Behmke, M.~Cerchez,
  O.~Jäckel, D.~Hemmers, T.~Toncian, M.~C. Kaluza, A.~Belyanin, G.~Pretzler,
  O.~Willi, A.~Pukhov, M.~Zepf, and G.~G. Paulus.
\newblock Harmonic generation from relativistic plasma surfaces in ultrasteep
  plasma density gradients.
\newblock {\em Physical Review Letters}, 109(12):125002, 2012.

\bibitem{Kahaly2013}
S.~Kahaly, S.~Monchocé, H.~Vincenti, T.~Dzelzainis, B.~Dromey, M.~Zepf, Ph.~Martin, F.~Quéré.
\newblock Direct Observation of Density-Gradient Effects in Harmonic Generation from Plasma Mirrors.
\newblock {\em Physical Review Letters}, 110(17), 2013.

\bibitem{Dromey2009}
B.~Dromey, D.~Adams, R.~Hörlein, Y.~Nomura, S.~G. Rykovanov, D.~C. Carroll,
  P.~S. Foster, S.~Kar, K.~Markey, P.~McKenna, D.~Neely, M.~Geissler, G.~D.
  Tsakiris, and M.~Zepf.
\newblock Diffraction-limited performance and focusing of high harmonics from
  relativistic plasmas.
\newblock {\em Nature Physics}, 5(2):146--152, 2009.

\bibitem{Bierbach2012}
J.~Bierbach, C.~Rödel, M.~Yeung, B.~Dromey, T.~Hahn, A.~G. Pour, S.~Fuchs,
  A.~E. Paz, S.~Herzer, S.~Kuschel, O.~Jäckel, M.~C. Kaluza, G.~Pretzler,
  M.~Zepf, and G.~G. Paulus.
\newblock Generation of 10 $\mu$w relativistic surface high-harmonic radiation
  at a repetition rate of 10 hz.
\newblock {\em New Journal of Physics}, 14, 2012.

\bibitem{Watts2002}
I.~Watts, M.~Zepf, E.~L. Clark, M.~Tatarakis, K.~Krushelnick, A.~E. Dangor,
  R.~M. Allott, R.~J. Clarke, D.~Neely, and P.~A. Norreys.
\newblock Dynamics of the critical surface in high-intensity laser-solid
  interactions: Modulation of the xuv harmonic spectra.
\newblock {\em Physical Review Letters}, 88(15), 2002.

\bibitem{Teubner2003}
U.~Teubner, G.~Pretzler, T.~Schlegel, K.~Eidmann, E.~Förster, and K.~Witte.
\newblock Anomalies in high-order harmonic generation at relativistic
  intensities.
\newblock {\em Physical Review A}, 67(1), 2003.

%\bibitem{Boyd2007}
%T.~J.~M. Boyd and R.~Ondarza-Rovira.
%\newblock Plasma modulation of harmonic emission spectra from laser-plasma
  interactions.
%\newblock {\em Physical Review Letters}, 98(10), 2007.

\bibitem{Sandberg2007}
R.~L. Sandberg, A.~Paul, D.~A. Raymondson, S.~Hadrich, D.~M. Gaudiosi,
  J.~Holtsnider, R.~I. Tobey, O.~Cohen, M.~M. Murnane, H.~C. Kapteyn, C.~G.
  Song, J.~W. Miao, Y.~W. Liu, and F.~Salmassi.
\newblock Lensless diffractive imaging using tabletop coherent high-harmonic
  soft-x-ray beams.
\newblock {\em Physical Review Letters}, 99(9), 2007.

\bibitem{Lambert2008}
G.~Lambert, T.~Hara, D.~Garzella, T.~Tanikawa, M.~Labat, B.~Carre, H.~Kitamura,
  T.~Shintake, M.~Bougeard, S.~Inoue, Y.~Tanaka, P.~Salieres, H.~Merdji,
  O.~Chubar, O.~Gobert, K.~Tahara, and M.~E. Couprie.
\newblock Injection of harmonics generated in gas in a free-electron laser
  providing intense and coherent extreme-ultraviolet light.
\newblock {\em Nature Physics}, 4(4):296--300, 2008.

\bibitem{Quere2006}
F.~Quéré, C.~Thaury, P.~Monot, S.~Dobosz, P.~Martin, J.~P. Geindre, and
  P.~Audebert.
\newblock Coherent wake emission of high-order harmonics from overdense
  plasmas.
\newblock {\em Physical Review Letters}, 96(12), 2006.

\bibitem{Dromey2009a}
B.~Dromey, S.~G. Rykovanov, D.~Adams, R.~Hörlein, Y.~Nomura, D.~C. Carroll,
  P.~S. Foster, S.~Kar, K.~Markey, P.~McKenna, D.~Neely, M.~Geissler, G.~D.
  Tsakiris, and M.~Zepf.
\newblock Tunable enhancement of high harmonic emission from laser solid
  interactions.
\newblock {\em Physical Review Letters}, 102(22), 2009.

\bibitem{Fidone1978}
I.~Fidone, G.~Ramponi, and P.~Brossier.
\newblock Nonthermal emission at plasma frequency.
\newblock {\em Physics of Fluids}, 21(2):237--238, 1978.

\bibitem{Forsyth1949}
P.~A. Forsyth, W.~Petrie, and B.~W. Currie.
\newblock Auroral radiation in the 3,000-megacycle region.
\newblock {\em Nature}, 164(4167):453--453, 1949.

\bibitem{Wild1953}
J.~P. Wild, J.~D. Murray, and W.~C. Rowe.
\newblock Evidence of harmonics in the spectrum of a solar radio outburst.
\newblock {\em Nature}, 172(4377):533--534, 1953.

\bibitem{Teubner1999}
U.~Teubner, P.~Gibbon, D.~Altenbernd, D.~Oberschmidt, E.~Forster,
  A.~Mysyrowicz, P.~Audebert, J.~P. Geindre, and J.~C. Gauthier.
\newblock Plasma frequency and harmonic emission from fs-laser plasmas.
\newblock {\em Laser and Particle Beams}, 17(4):613--619, 1999.

\bibitem{Boyd2000}
T.~J.~M. Boyd and R.~Ondarza-Rovira.
\newblock Plasma line emission from short pulse laser interactions with dense
  plasmas.
\newblock {\em Physical Review Letters}, 85(7):1440--1443, 2000.

\bibitem{Kunzl2003}
T.~Kunzl, R.~Lichters, and J.~Meyer-Ter-Vehn.
\newblock Large-amplitude plasma waves and 2 omega(p) emission driven by
  laser-generated electron jets in overdense plasma layers.
\newblock {\em Laser and Particle Beams}, 21(4):583--591, 2003.

\bibitem{comment}
The plasma resonance has a considerable width due to damping. As it is excited by the laser frequency $\omega_L$, the dominant mode will have a frequency equal to an integer multiple of $\omega_L$.

\bibitem{Brunel1987}
F.~Brunel.
\newblock Not-so-resonant, resonant absorption.
\newblock {\em Physical Review Letters}, 59(1):52--55, 1987.

\bibitem{Geindre2010}
J.~P. Geindre, R.~S. Marjoribanks, and P.~Audebert.
\newblock Electron vacuum acceleration in a regime beyond brunel absorption.
\newblock {\em Physical Review Letters}, 104(13), 2010.


\end{thebibliography}
\end{document}